\documentclass[12pt]{report}
\usepackage[utf8]{inputenc}

\usepackage[paper=letterpaper,left=1in,right=1in,top=1in,bottom=1in]{geometry}

\usepackage[T1]{fontenc}
\usepackage{fbb}

\usepackage[symbol]{footmisc}
\usepackage{graphicx}
\usepackage{sidecap}
\usepackage{setspace}
 
\usepackage[numbers,sort&compress]{natbib}

\newcommand{\FeK}{Fe K$\alpha$~}

\usepackage{xcolor}
\definecolor{darkred}{rgb}{0.8,0.0,0.0}
\definecolor{darkblue}{rgb}{0.0,0.0,0.5}

\usepackage{hyperref}
\hypersetup{
    colorlinks = true,
    citecolor = darkred
}

\begin{document}
\begin{titlepage}
	\centering
	\vspace{1cm}
	\rule{\textwidth}{0.4pt}
	\vspace{0.3cm}
	{\huge\bfseries Supermassive Black Hole Spin and Reverberation\par}
	\vspace{-0.3cm}
	\rule{\textwidth}{0.4pt}
	\vspace{0.5cm}
	{\large White Paper Submitted to Astro 2020 Decadal Survey\\
	        Thematic Science Area:\\
	        Primary: \emph{Galaxy Evolution}\\
	        Secondary: \emph{Properties of Supermassive Black Holes, Active Galactic Nuclei.}\par}
	\vspace{2cm}
	{\large\itshape A. Zoghbi\footnote{abzoghbi@umich.edu}, D. R. Wilkins, L. Brenneman, G. Miniutti, G. Matt, J. Garcia, E. Kara, E. Cackett, B. De Marco, M. Dovciak, P. Tzanavaris, A. Hornschemeier, E. Bulbul, J. Miller, R. Kraft, A. Ptak, 
	R. Smith, R. Petre.}
	
	\vfill
	X-ray reverberation mapping has emerged as a powerful probe of microparsec scales around AGN, and with high sensitivity detectors, its full potential in echo-mapping the otherwise inaccessible disk-corona at the black hole horizon scale will be revealed.

	\vfill

\end{titlepage}

\newpage
\subsection*{Introduction}



Studies of astrophysical Black Holes (BHs) are important for two main reasons. First, BHs represent an extreme manifestation of General Relativity (GR). Their immediate environments, along with neutron stars, are the only places in the Universe where the extraordinary properties of gravity in the strong limit are revealed. Second, the observed correlations between BH masses and the masses and binding energies of the bulges where they reside \cite{2000ApJ...539L...9F,2009ApJ...698..198G}, indicate that BHs, despite being small in size, play a crucial role in building and shaping galaxies \cite{2012ARA&A..50..455F}.

Accretion onto BHs can be very efficient, converting up to half of the accretion energy into radiation. Most of it is emitted in X-rays from within tens of gravitational radii ($r_g = GM/c^2\sim 1.5M_{\rm BH}$ km), where frame-dragging and gravitational redshifts become significant, and stable circular orbits disappear.
Direct imaging of these scales is currently possible only for Sgr~A$^\star$ and M~87, which are massive and close enough to be imaged with the event horizon telescope \citep{2011ApJ...738...38B}. At smaller angular scales, X-ray spectral and variability signatures provide the only indirect probes of these otherwise inaccessible regimes. 

While BH mass can be measured by its effect on parsec-scale gas, observations on the microparsec scale gas are needed for the spin. Yet, the spin is critical in understanding the BH formation history, and it may be responsible for the most powerful ejection phenomena.

The spin distribution of supermassive black holes (SMBH) may reveal their formation history \cite{1996MNRAS.283..854M,2008MNRAS.385.1621K,2013ApJ...775...94V}. BHs tend to be spun up if they grow primarily through accretion, unlike growth by randomly-oriented merger events. Efficiently-accreted gas flows in circular orbits down to the innermost stable circular orbit (ISCO), beyond which it plunges onto the BH. The ISCO depends monotonically on the spin, and therefore, the inner extent of the accretion disk gives a direct spin measure. For SMBH, modeling the relativistic iron line (Fe K$\alpha$) in X-rays has been the most reliable spin estimator. The low-energy extent of the line is shaped by gravitational redshift, and is therefore sensitive to the location of the ISCO \cite{2007ARA&A..45..441M}. The spins of about 30 AGN have been measured \cite{2013mams.book.....B,2016ASSL..440...99M} using this method so far. Most of them appear to be spinning close to the prograde maximum, but it is not clear if this is a true distribution or a result of modeling assumptions and observational biases. The next step requires higher sensitivity and energy resolution, which will not only increase the number of sources with spin measurements, but also allow us to better understand the modeling uncertainties (e.g. inclination and ionization) and absorption and emission originating in matter more distant from the black hole.

The field of BH astrophysics has witnessed a substantial transformation in recent years. Gravitational waves have been detected for the first time \cite{2016PhRvL.116f1102A}. 
The event GW170817 \cite{2017ApJ...848L..12A} specifically signaled the beginning of a new era of multi-messenger astronomy. Development of X-ray polarimetry has reached an advanced stage with IXPE \cite{2016SPIE.9905E..17W} and eXTP \cite{2016SPIE.9905E..1QZ}, and the next decade promises to allow observations of accreting BHs though a new and unexplored window \cite{2012MNRAS.426L.101M,2017ApJ...850...14B}. 
Theoretically, improvement in computational power and numerical techniques are accelerating the development of GR-MHD simulations of accretion disks around BHs that include radiation \cite{2014ApJS..213....7J,2018ApJ...857....1F}. Additionally, Time-domain astronomy facilities promise to accelerate the discovery of transient phenomena, which in many cases are driven by BH activity (e.g. tidal disruption events and quasar variability). All these developments necessitate a better understanding of how accretion at the BH scale proceeds.


\subsection*{X-ray Reverberation: State Of The Art}
\begin{figure}
    \centering
    \begin{tabular}{lcr}
        \includegraphics[height=110pt]{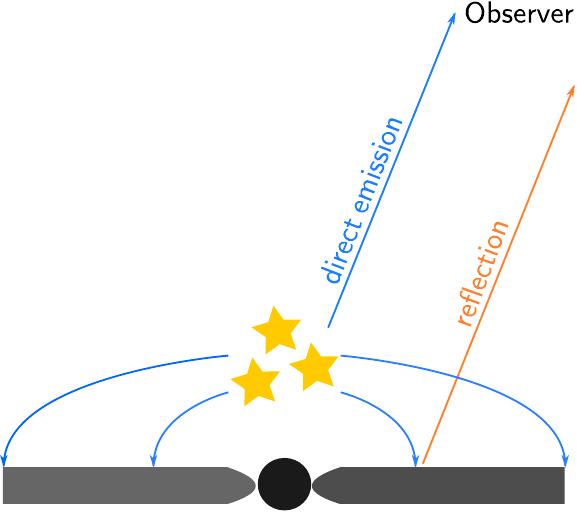} &
        \includegraphics[height=110pt]{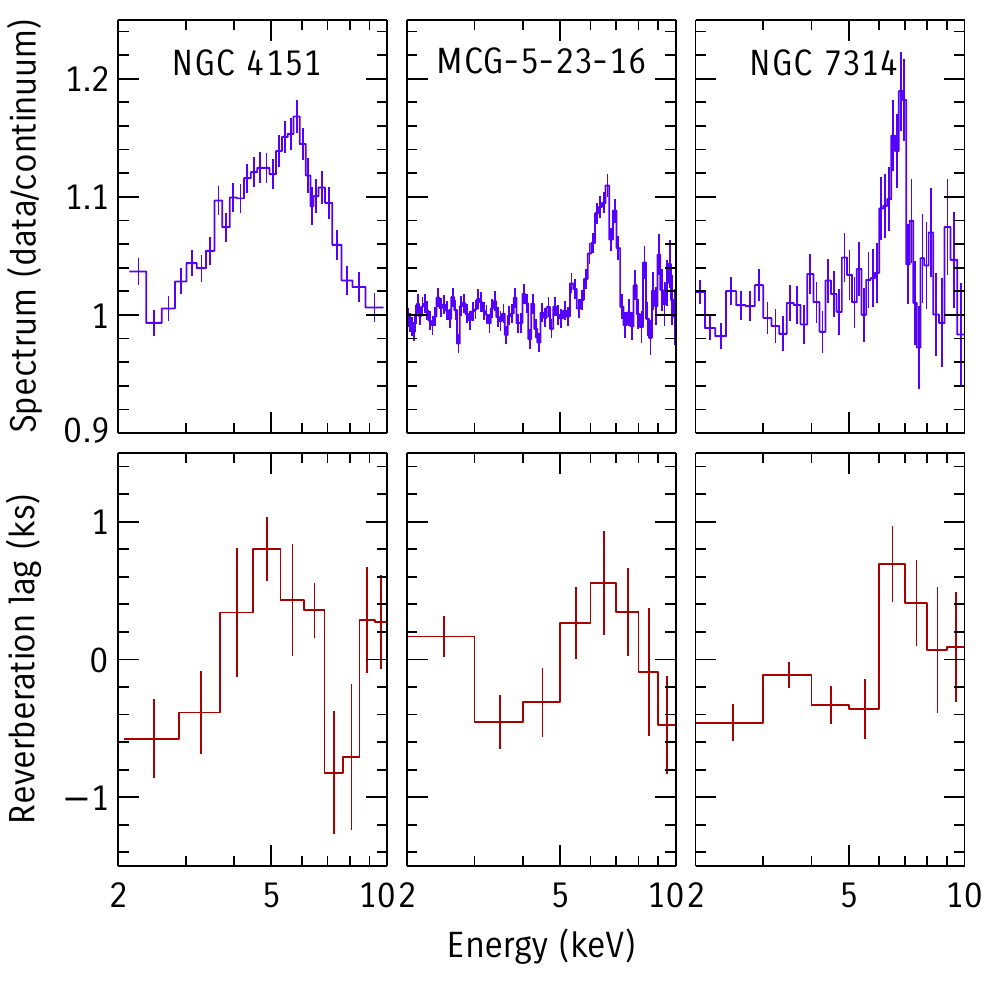} &
        \includegraphics[height=110pt]{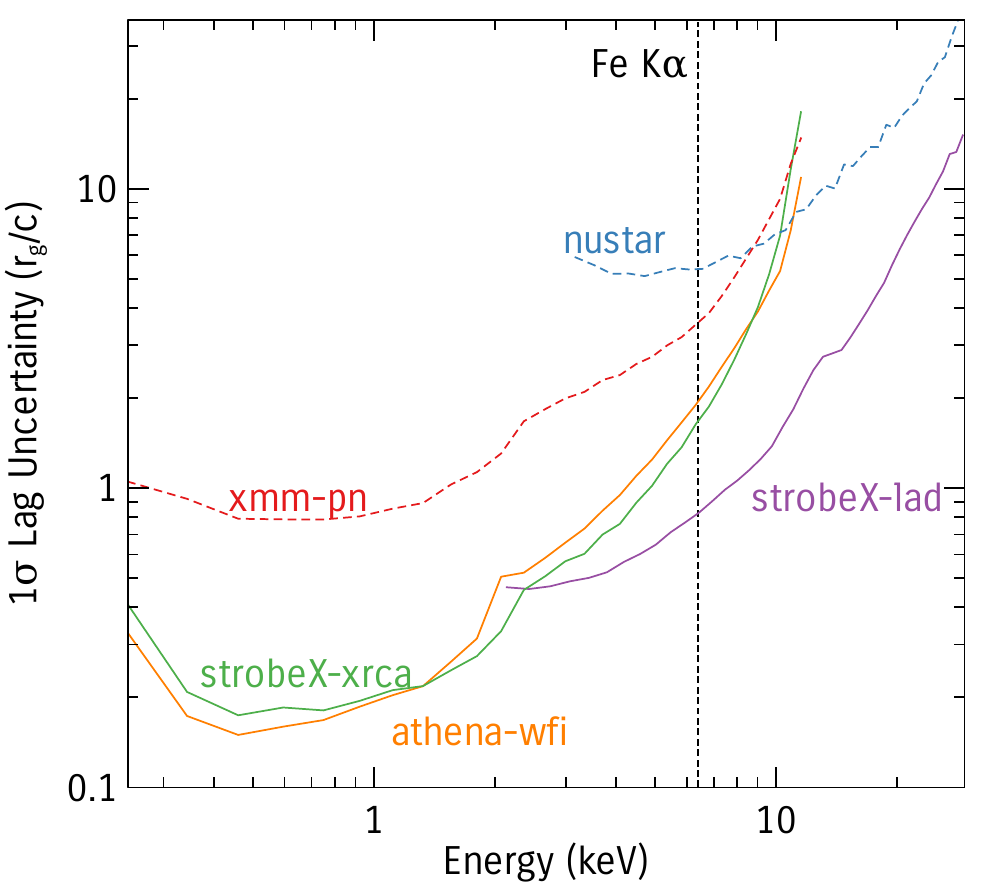}
    \end{tabular}
    \caption{\footnotesize {\em Left:} An illustration of how a hard X-ray source (corona) illuminating an accretion disk leads to reverberation, allowing the regions arround the BH to be mapped. {\em Middle:} A sample of the observed \FeK reverberation lags. The top panels show the spectra plotted as a ratio to the continuum, showing the relativistic \FeK line. The bottom panels show the measured delays from {\em XMM-Newton} showing how the peak of the line is delayed with respect to the continuum.
    {\em Right:} Lag sensitivity plot comparing current to future telescopes.}
    \label{fig:sensitivity:spin}
\end{figure}

The relativistic reflection spectrum is produced when a hard X-ray corona illuminates the dense accretion disk. The reflected spectral shape depends on the disk ionization, abundance and inclination and BH spin, with the Fe K$\alpha$ being the most prominent emission line \cite{2005MNRAS.358..211R,2013ApJ...768..146G}. Relativistic spectral signatures have been observed extensively over the last two decades \cite{2007ARA&A..45..441M,2014ApJ...788...76W}. It was realized early on \cite{1989MNRAS.238..729F,1990Natur.344..747S} that the physical separation of the corona and the reflecting accretion disk should produce a delay in the variations of the two components (Figure \ref{fig:sensitivity:spin}-left). The discovery had to wait until telescopes with large enough collecting area were operational. 

First, it was the soft excess ($<1$ keV), where the telescopes are most sensitive, that was observed to lag the hard continuum \cite{2009Natur.459..540F,2010MNRAS.401.2419Z}. Delays in the \FeK line were later observed too \cite{2012MNRAS.422..129Z}. A survey of the brightest Seyfert galaxies indicates that about half show \FeK line reverberation \citep{2016MNRAS.462..511K}. The time delays range between 10-1000s of seconds, corresponding to light travel distances less than $\sim 10~r_g$ for $M_{\rm BH}\sim10^{6}-10^{8} M_{\odot}$. The detection of the \FeK delays (Figure \ref{fig:sensitivity:spin}-middle) opens a new window of discovery, first because the \FeK is relatively clean compared to the soft excess. Second, the delays provided unprecedented physical size measures at the BH scale that is independent of the spectral modeling \cite{2014A&ARv..22...72U}.

The magnitudes of the lags scale with BH mass \cite{2013MNRAS.431.2441D,2016MNRAS.462..511K}, implying that the emission regions in different sources are comparable when measured in $r_g$ ($\propto M_{\rm BH}$).
The delays also provide evidence that the corona is {\em compact}, not just physically, but also radiatively. The short delays and coronal temperatures measured with spectral curvatures in hard X-rays with {\em NuSTAR} \cite{2013ApJ...770..103H}, showed that pair production and annihilation are essential ingredients in the corona, controlling the shape of the observed spectra \cite{2015MNRAS.451.4375F,2018ApJ...866..124K}.

A major unknown in modeling the shape of the relativistic spectra is the disk-corona geometry. A lamp post is often assumed, where the corona is assumed to be a point on the spin axis of the BH. Alternative modeling attempts have been made to recover the illumination pattern on the disk directly from the data assuming broken powerlaw form. A recent breakthrough however, allowed the profile to be measured directly from high quality spectra \cite{2012MNRAS.424.1284W}. Measuring the {\em emissivity profile} and its variations over time have been used to infer changes in the geometry during and after flaring states in an unprecedented manner \cite{2015MNRAS.449..129W}.

In a few cases, reverberation studies have moved beyond a simple delay measurement. Variations in the delay as a function of flux \cite{2013MNRAS.430.1408K} and/or time \cite{2017MNRAS.471.4436W} have revealed that {\em the disk-corona geometry is very dynamic}, as expected in a turbulent flow. When the source is bright, the corona is observed sometimes to be horizontally extended above the disk to scales of a few tens of $r_g$, and when the flux drops, it is compact (a few $r_g$), with some evidence of vertical structure, reminiscent of collimated jet outflows. The higher sensitivity of future instruments will provide the critical signal to probe the dynamic corona in unprecedented details.


\subsection*{The Next Decade}
The next generation of telescopes ({\em Athena, STROBE-X}) are expected to produce unprecedented quality spectra within short exposures. This will not just increase the number of spin measurements to span different BH mass and Eddington ratios. Time-resolved spectral modeling, where parameters such as spin are not expected to change, while others related to geometry, ionization and illumination may change, will be crucial in understanding the systematic uncertainties in currents spin measurements (see white paper by Garcia et al.)

Just as the large effective area of XMM-Newton and NuSTAR in the \FeK band has been instrumental in the discovery of relativistic reverberation, the area increase with future instruments will allow delays that are a fraction of the light-crossing time at the event horizon for a large number of bright active nuclei to become accessible (Figure \ref{fig:sensitivity:spin}-right). Currently, $\sim30$ sources have measured reverberation lags in the soft band, but with future telescopes, the number will roughly triple, including higher redshifts sources and a wider range of accretion rates. Current reverberation measurements, which in many cases amount only to a characteristic time delay giving a size scale, demonstrate the efficiency of the techniques. The full potential of reverberation mapping awaits the telescopes of the next decade. In the following, we discuss some of the crucial questions that reverberation studies will tackle.

The corona is extremely hot ($\sim10^9$ degrees) and can produce up to $\sim50\%$ of the bolometric luminosity of the AGN, and yet, what powers this corona is unknown. It may be related to magnetic reconnection or jet launching. Understanding the geometry of the corona can put constraints on these models. The theoretical relativistic response of an accretion disk can be calculated using ray-tracing simulations, and it depends on both time and energy. For a point corona illuminating a flat disk, the response has been studied extensively \cite{1995MNRAS.272..585C,1999ApJ...514..164R,2014MNRAS.439.3931E,2014MNRAS.438.2980C}. In this simple case, the response is most sensitive to the distance between the corona and the reflector, and it is sufficient in modeling most of current measurements.

\begin{figure}
    \centering
    \includegraphics[height=120pt]{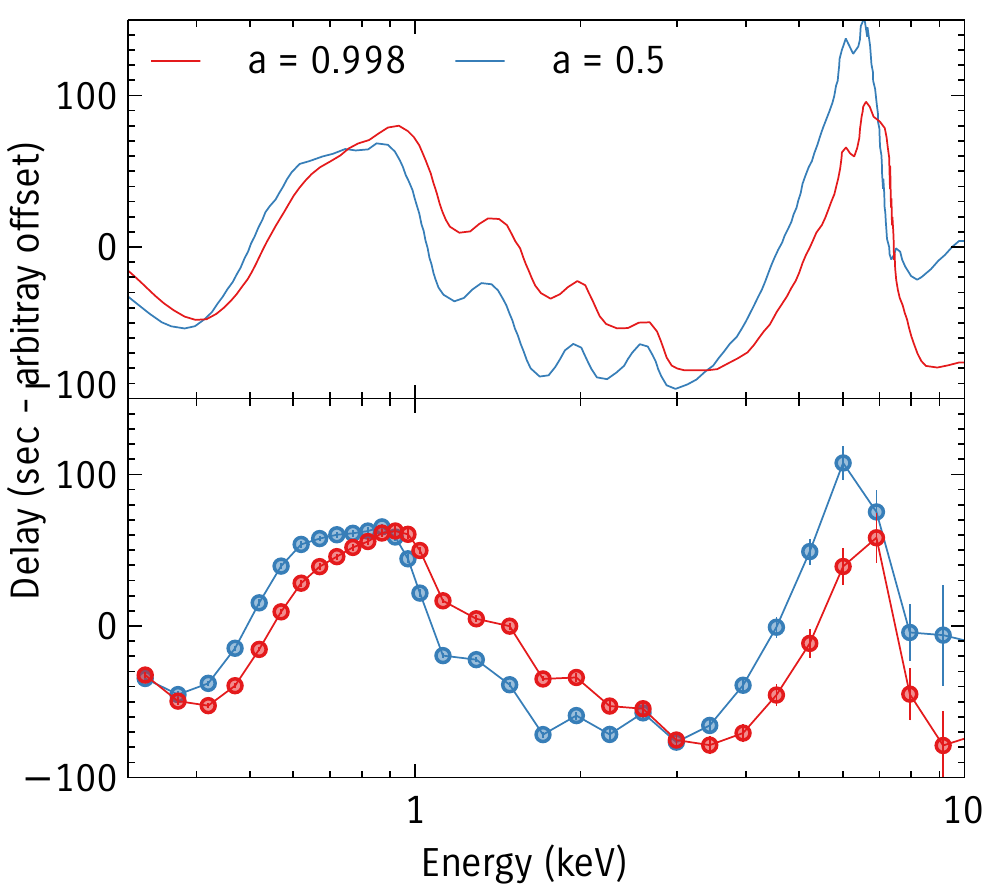}
    \includegraphics[height=120pt]{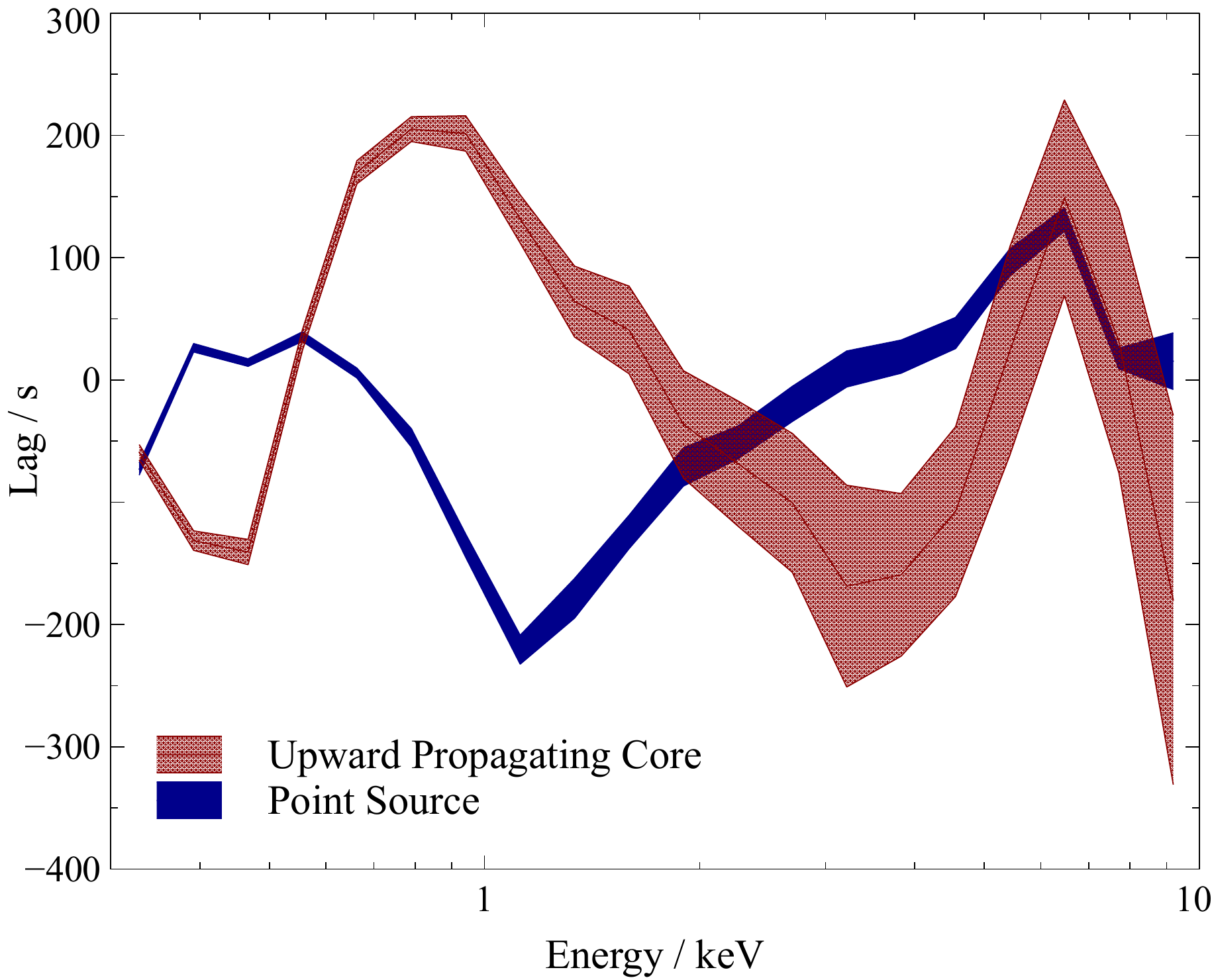}
    \includegraphics[height=120pt]{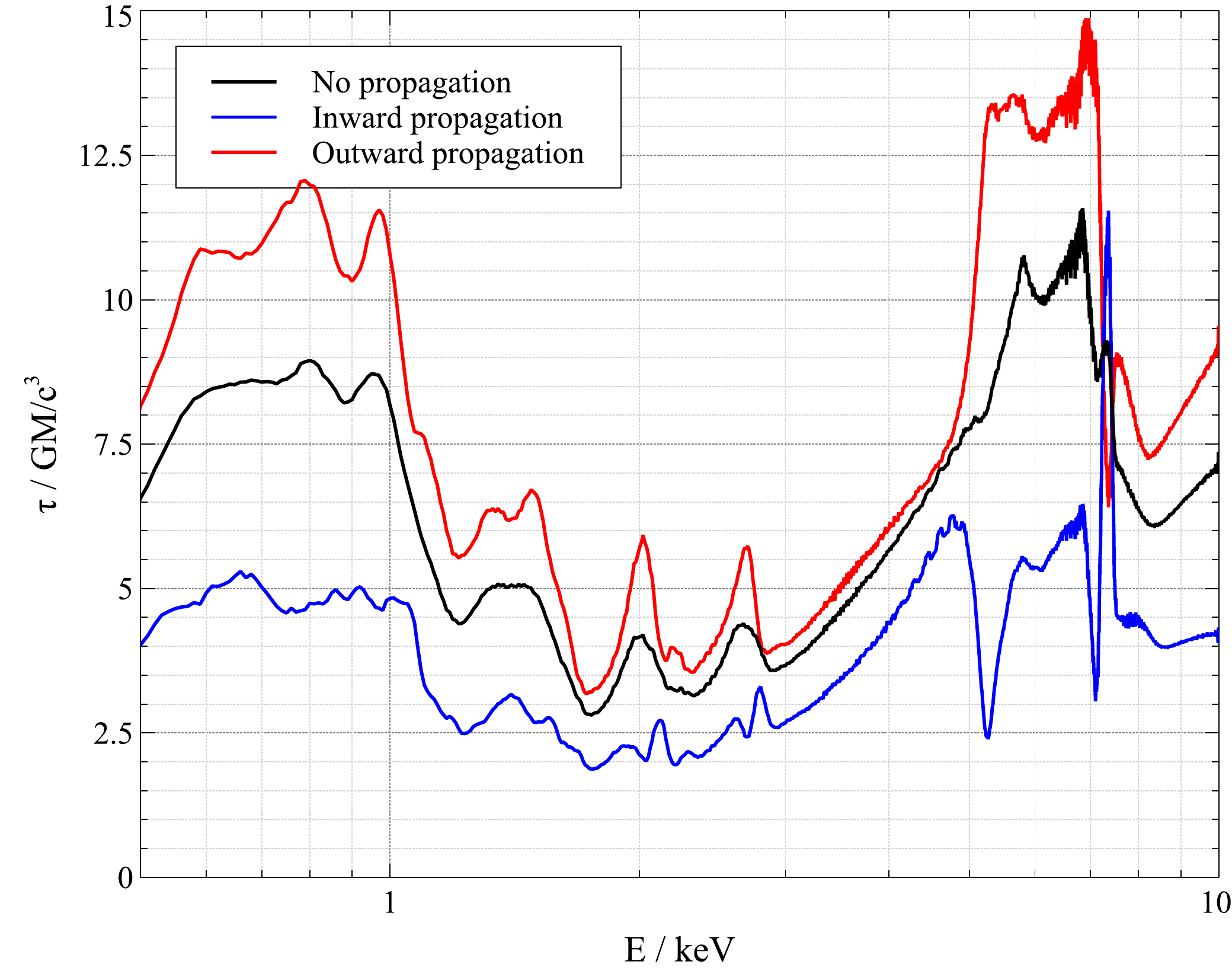}
    \caption{\footnotesize {\em Left-Top}: The lag-energy spectrum represented by the average arrival time of photons as a function of energy for two BH spins. Luminosity fluctuations propagate up a vertically collimated corona extending 10 rg above the plane of the disc at $0.01 c$ \citep[from][]{2016MNRAS.458..200W}.
    {\em Left-Bottom:} A simulated lag-energy spectrum showing the quality of data predicted from the next generation of detectors (WFI on Athena or XRCA on STROBE-X; 100 ks exposure for $M=10^{6}M_{\odot}$). Notice that the earliest arriving photons (the most negative) change as the BH spin changes. \textit{Middle:} Simulated lag-energy spectra for a 500ks observation of a typical bright Seyfert galaxy with $M_\mathrm{BH} = 10^7M_\odot$ with the \textit{Athena WFI}, is a sensitive probe of the geometry of the corona and how luminosity fluctuations propagate through its extent. The lag-energy spectrum arising from a point source can be readily distinguished from a collimated core, embedded within the central regions of an extended corona, through which fluctuations propagate upwards. \textit{Right: } Lag-energy spectra from different models for the flow within the corona. The models give different signatures that can be probed with instruments of the next decade.}
    \label{fig:geometry}
\end{figure}

In practice, the corona is likely extended. Current high quality data already provide some evidence for this \cite{2017MNRAS.471.4436W}, with the slow variability influenced by an extended region, while a compact core dominates the fastest variability. Different coronal geometries illuminate the disk differently, predicting different signatures in the reverberation response (Figure \ref{fig:geometry}). {\em Measurements of how the reflected emission responds to continuum variations, the geometry can be inferred}. A compact or vertically extended corona might be associated with the base of a jet, while a corona that extends over the disk is likely associated with disk instabilities. 

The quest for the system geometry also includes {\em the geometry of the reflector itself}, analogous to the geometry of the Broad Line Region in optical reverberation studies \cite{2013ApJ...764...47G}. Internal pressures within the accretion disk would naturally result in nonzero scale heights, which may not be negligible compared to the size of the corona, especially in super-Eddington flows where the radiative efficiency is expected to be small. Additionally, misalignment between the outer gas flowing inward and the BH spin (and the inner disk) is expected to be common in randomly oriented accretion events from the galaxy \cite{1992MNRAS.258..811P,2005MNRAS.363...49K}. Such deviations from a thin disk produce measurable effects on the reverberation response \cite{2018ApJ...868..109T}. This will not be just a powerful tool to probe these likely common features, but also a {\em tool for studying super-Eddington transient events} such as those from the BH disruption of stars \cite{2016Natur.535..388K}.

\begin{figure}
    \centering
    \includegraphics[height=110pt]{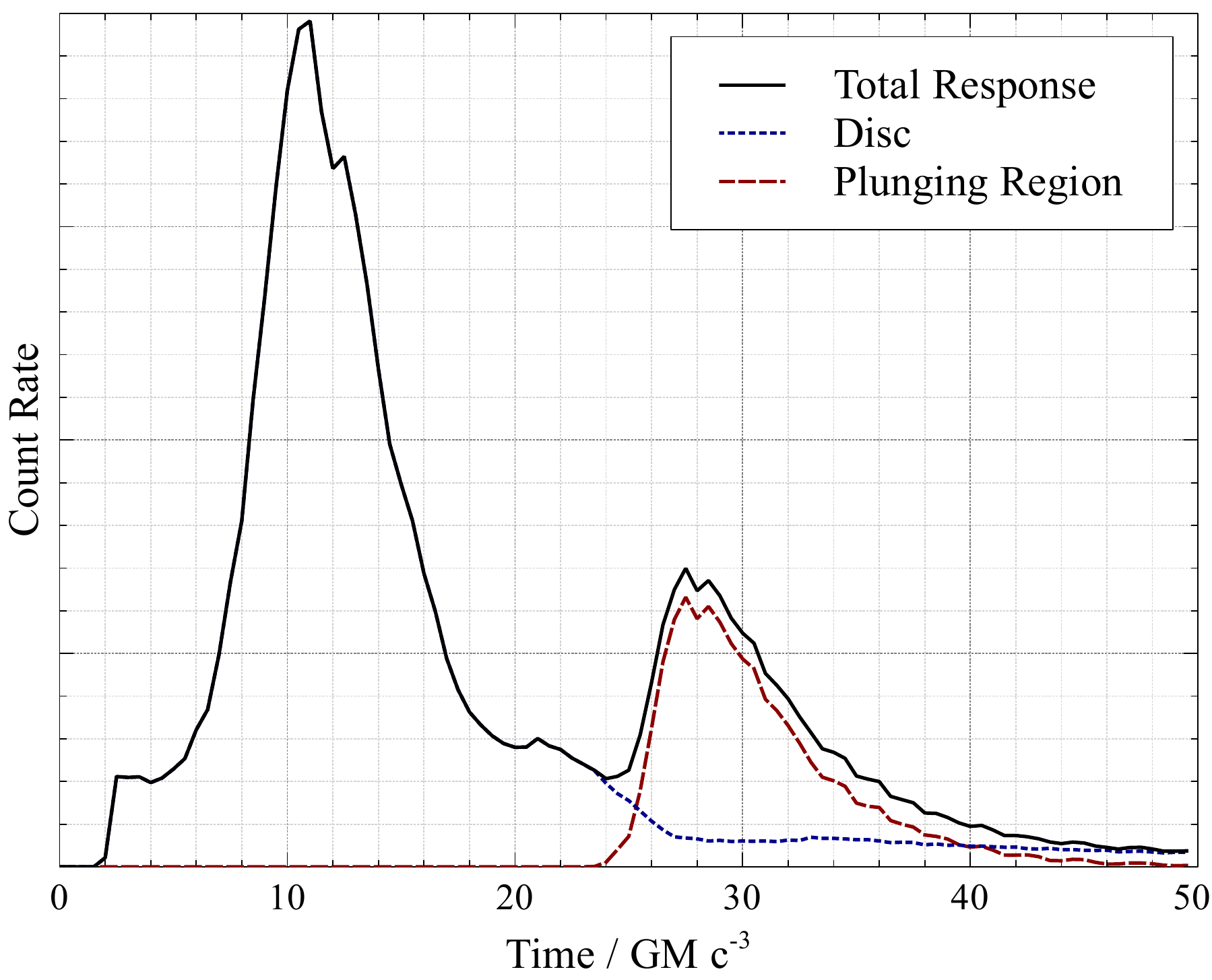}
    \includegraphics[height=110pt]{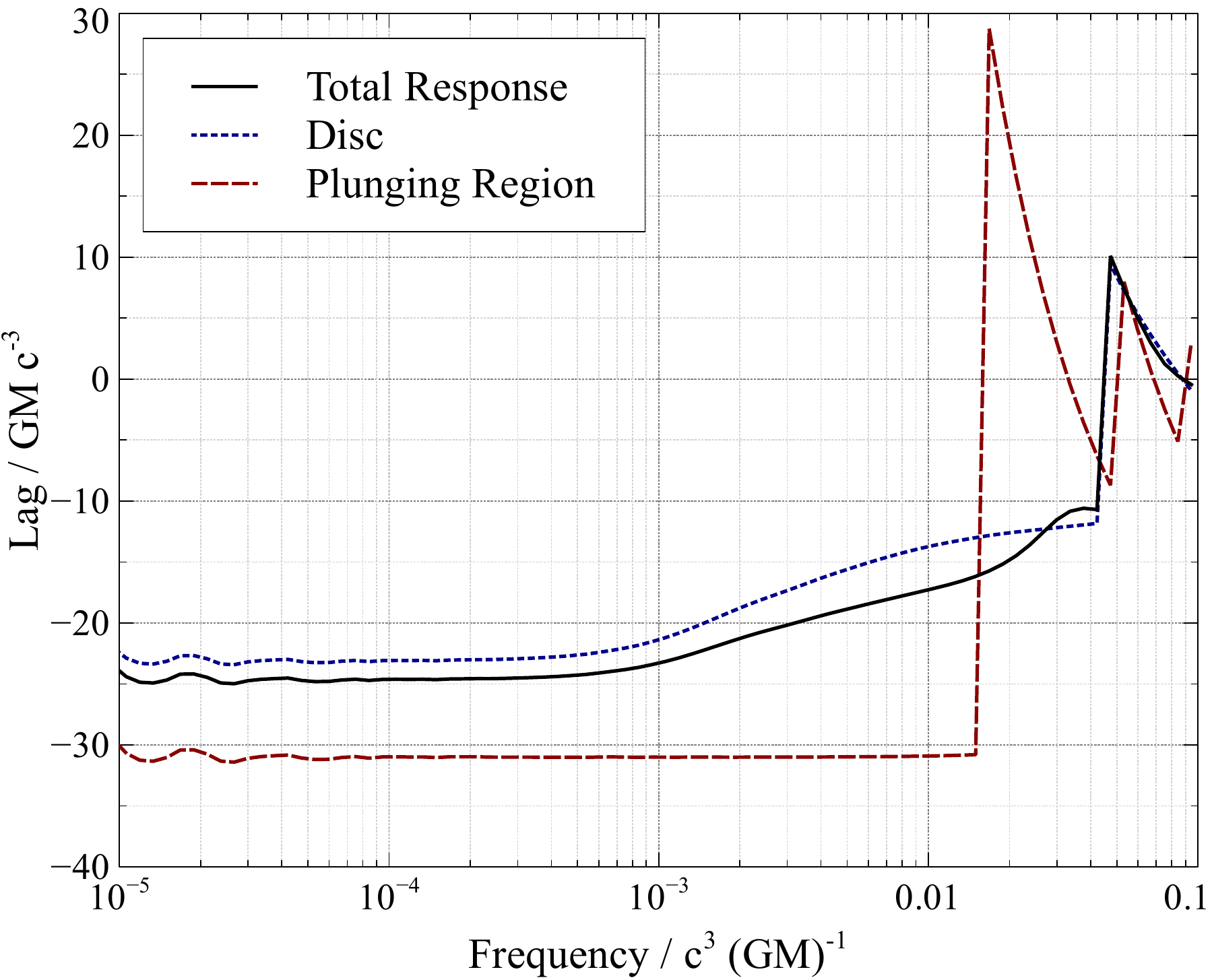}
    \includegraphics[height=110pt]{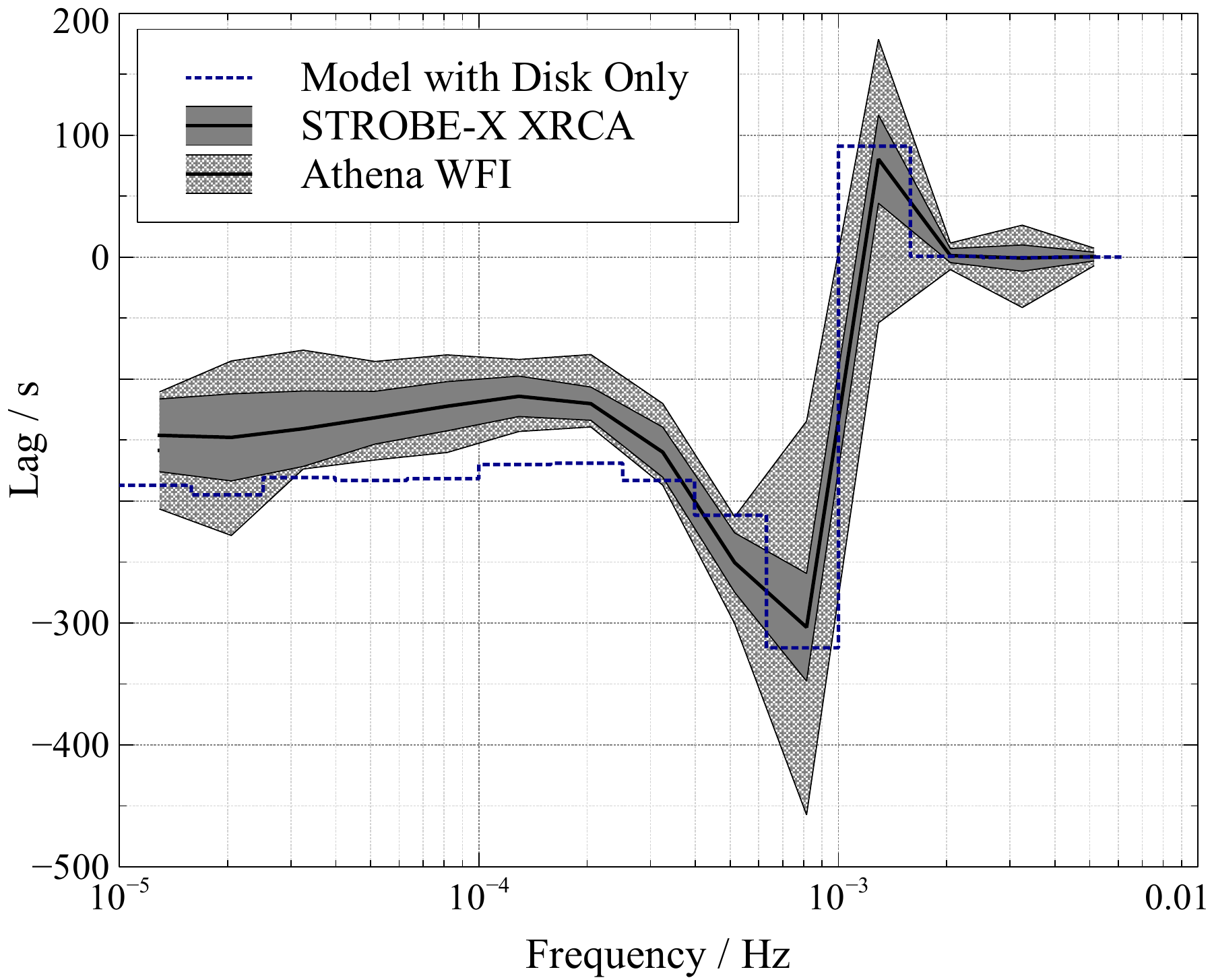}
    \caption{\footnotesize Reverberation from the plunging region. {\em Left:} X-ray reverberation from the innermost parts of the flow where material plunges from the innermost stable orbit to the event horizon produces a secondary response of delayed, extremely redshifted \FeK emission from highly ionized material on ballistic orbits around a maximally spinning BH (a=0.998). The lags are for the 1--2 keV band. X-ray reverberation from the innermost parts of the flow where material plunges from the innermost stable orbit to the event horizon produces additional delayed redshifted emission compared to just the accretion disc that can be detected in the highest frequency components of the variability. Probing the dynamics of the plunging region provides a unique probe of the strong gravity and extreme environment close to the event horizon of the BH. {\em Middle:} The lag as a function of Fourier frequency for the 1-2keV band in which extremely redshifted \FeK line emission from the innermost regions of the accretion flow around a maximally spinning (a=0.998M) BH. Reverberation from the plunging region imprints a detectable signature on the shape of the lag-frequency spectra. \textit{Right:} Simulated lag-frequency spectra comparing the 1-2keV band with the 0.3-1keV band using the next-generation X-ray missions \textit{Athena} and \textit{STROBE-X}. In the case of bright AGN, it will be possible to distinguish the reverberation from the plunging region from the signal detected from the disk alone.}
    \label{fig:plunging}
\end{figure}

High precision spectral-timing measurements can also open the door for exploring other aspects of BHs, potentially allowing for strong field tests for GR. The no-hair theorem, which states that astrophysical BH can be described only by their mass and spin, yields specific predictions about the shape of the iron line, which are testable with high quality spectra \cite{2013ApJ...773...57J}.
More recent work has additionally explored the reverberation signatures from the plunging region beyond the ISCO \citep[Figure \ref{fig:plunging};][]{plunging}. The quest is to identify the observable signatures from such a region, and use observations to distinguish a plunging region from a no-ISCO case.

In addition to the geometry, reverberation mapping is also sensitive to other parameters of the system such as the {\em BH spin}. Models of propagating coronal flow indicate that the energies of the earliest arriving photons in the spectrum provide depends strongly on the spin. A 3 keV dip is expected in the lag-energy spectra (Figure \ref{fig:geometry}-left) only in the presence of strong gravity and when the accretion disc extends inwards within 5 $r_g$. The feature becomes weaker for more slowly spinning BHs or where the accretion disc is truncated, and it is not seen for spin parameters $a < 0.5$. These signatures provide independent consistency checks to spectroscopic measurements from the red wing of the \FeK line. Reverberation lags also depend on the {\em BH mass}. Simple modeling of the observed delays show that masses inferred from relativistic reverberation modeling are consistent with those measured from other methods \cite{2014MNRAS.439.3931E}. The larger reverberation samples, combined with more precise measurements expected from {\em Athena} and {\em STROBE-X}, will provide new BH mass estimates. High-sensitivity telescopes with monitoring capabilities will extend the reverberation studies to higher mass SMBHs as the delays are longer. This capability will also allow reverberation studies of the narrow \FeK line \cite{narrow_line}, which is crucial for mapping the sub-parsec circum-nuclear structure of these SMBHs.

\bibliographystyle{unsrtnat}
\subsection*{References}
{\def\section*#1{}\bibliography{main}}

\end{document}